\documentclass[useAMS,usenatbib]{mn2e}

\usepackage{fixltx2e} 
\usepackage{graphicx}
\usepackage{color,ulem}
\usepackage{dcolumn}
\usepackage{bm}
\usepackage{amsmath,amssymb}
\usepackage[T1]{fontenc}
\usepackage{aecompl}

\definecolor{webgreen}{rgb}{0,.5,0}
\definecolor{webbrown}{rgb}{.6,0,0}
\definecolor{purple}{rgb}{0.5,0,.5}

\newcommand{\mx}{_\mathrm{m}}
\newcommand{\trunc}{_\mathrm{t}}
\newcommand{\infl}{_\mathrm{i}}

\newcommand{\kpc}{\,{\rm kpc}}

\title{3D morphology of a random field from its 2D cross-section}

\author[I.~Makarenko, A.~Fletcher, \& A.~Shukurov]{Irina Makarenko,
Andrew Fletcher, \& Anvar Shukurov\thanks{E-mail: irina.makarenko@ncl.ac.uk;
andrew.fletcher@ncl.ac.uk; anvar.shukurov@ncl.ac.uk}\\
School of Mathematics and Statistics, Newcastle University,
Newcastle upon Tyne, NE1~7RU, UK}

\begin{document}
\pagerange{\pageref{firstpage}--\pageref{lastpage}} \pubyear{2014}

\maketitle

\label{firstpage}

\begin{abstract}
The two aspect ratios of randomly oriented triaxial ellipsoids (representing
isosurfaces of an isotropic 3D random field) can be determined from a single
2D cross-section of their sample using the probability density function (PDF)
of the \textit{filamentarity\/} $F$ of individual structures seen in
cross-section ($F=0$ for a circle and $F=1$ for a line). The PDF of $F$ has a
robust form with a sharp maximum and truncation at larger $F$, and the most
probable and maximum values of $F$ are uniquely and simply related to the two
aspect ratios of the triaxial ellipsoids. The parameters of triaxial
ellipsoids of randomly distributed sizes can still be recovered from the PDF
of $F$. This method is applicable to many shape recognition problems, here
illustrated by the neutral hydrogen density in the turbulent interstellar
medium of the Milky Way. The gas distribution is shown to be filamentary with
axis ratios of about 1:2:20.
\end{abstract}
\begin{keywords}
turbulence -- methods: data analysis -- methods: statistical -- ISM: clouds
\end{keywords}
\maketitle

\section{Introduction}An efficient approach to quantify the morphology of a
random field is based on the classification of its excursion sets
\citep{AT07}, i.e., the interior of its isosurfaces. The isosurfaces of a
differentiable random field at a sufficiently high level can be approximated
by ellipsoids. Thus, we consider the morphology of structures in a spatial
cross-section of a system consisting of a large number of randomly oriented
structures (both infinitely long cylinders and triaxial ellipsoids) which
represent the isosurfaces of a random field. The approach we develop allows us
to retrieve the 3D morphology of the objects (i.e., their aspect ratios) from
a \textit{single\/} 2D spatial cross-section of the system.

The recovery of 3D shapes from a small number of their 2D cross-sections or
projections is a standard problem of \textit{stereology\/} (distinct from
tomography where a sufficiently large number of such 2D images is available)
\citep{SW08}. Recovering the intrinsic 3D morphology from a \textit{single\/}
2D image, which is a common problem in astronomy and cosmology, is a
challenging problem. Approaches based on the Fourier transform of a 2D
projection are limited by strong restrictions; in particular, all the objects
must have their symmetry axes aligned with the projection plane or at least
three projections are required \citep{Ry87}. Extensive surveys of the
distribution of galaxies, and their interpretation in terms of galaxy
formation theories, require statistical analysis of the morphology of a large
number of astronomical images \citep[e.g.,][]{VFNWG14}. A powerful method for
the statistical morphological analysis of convex objects in 3D using Minkowski
functionals, is provided by Blaschke diagrams \citep{SBMSSS99,RDP10a}, but we
are not aware of any applications of this approach to the deprojection problem
from 2D to 3D.

We employ the probability distribution of the \textit{filamentarity\/}
\citep{Mecke94,Sahni98}, a dimensionless morphological characteristic based on
the Minkowski functionals, of the structures seen in a single random  2D
cross-section through a sample of randomly oriented 3D objects, to derive
statistically the intrinsic, 3D morphological properties of the sample.
Unsurprisingly, this requires additional assumptions, but these are not
particularly restrictive in our approach. Here we assume only that the objects
are oriented isotropically and do not overlap in 3D. We consider 2D
cross-sections in this paper; the statistical properties of 2D projections and
extension to anisotropic distributions will be discussed elsewhere.

In 2D, the three Minkowski functionals of a closed contour are its enclosed
area $S$, perimeter $P$, and the Euler characteristic \citep{AT07}, but it is
convenient to introduce the dimensionless \textit{filamentarity:\/}
\begin{equation}
F = (P^2 - 4 \pi S)/(P^2 + 4 \pi S)\,.
\label{eq:F}
\end{equation}
By definition, $0 \leq F \leq 1$, with $F = 0$ for a circle, $F = (4 - \pi)/(4
+ \pi) \approx 0.12$ for a square and $F = 1$ for a line segment (not
necessarily straight). For an ellipse with semi-major and semi-minor axes $a$
and $b$, the area and perimeter can be calculated as
\begin{equation}\label{eq:SP}
S = \pi a b\,,
\quad P \approx \pi (a + b) \frac{1 - 3 \lambda^4/64}{1 - \lambda^2/4}\,,
\quad \lambda=\frac{a-b}{a+b}\,,
\end{equation}
with an accuracy better than $0.2\%$ for $P$. It is useful to write out a
relation between the filamentarity and the aspect ratio of a rectangle with an
aspect ratio $x=a/b$:
\begin{equation}\label{Fx}
F=[(x+1)^2-\pi x]/[(x+1)^2+\pi x]\,.
\end{equation}
A similar relation can be written for ellipses using Eq.~\eqref{eq:SP} but it
is less compact and we omit it here. Importantly, the Minkowski functionals
are sensitive to all statistical moments of the random field
\citep[e.g.,][]{WBO14}.

The filamentarity $F$ is only sensitive to the shape of a structure, but not
to its size. Therefore, this diagnostic is especially useful as a quantifier
of multi-scale, self-similar, intermittent structures, such as those in
turbulent flows. Filamentarity (and other similar quantities) has been used as
a \textit{shapefinder\/} in statistical analyses of the isodensity contours of
large-scale galaxy distributions \citep{Mecke94,Sahni98,SBMSSS99}. However, we
are not aware of any earlier exploration of the statistical properties of the
filamentarity itself. Apart from probability distributions of the Euler
characteristic, previous applications only involve the mean values of the
Minkowski functionals or related morphological measures
\citep{PB96,SW08,WBO14}. In contrast, here we consider the probabilistic
properties of the filamentarities of individual structures associated with a
random scalar field, such as gas density or temperature and emission or
absorption intensity distributions in a turbulent region.

\section{\label{model}The model}
To establish the statistical properties of the structures seen in a spatial
cross-section of an isosurface of a random field, we analyze a sample of the
cuts of a single ellipsoid by isotropically oriented and positioned planes
(called `cuts' to avoid confusion with the result of cutting an ensemble of
random structures in 3D, here called a `cross-section'). The ellipsoid
dimensions (principal axes) are refereed to as the thickness $T = \min(T, W,
L)$, width $W = {\rm med}(T, W, L)$ and length $L = \max(T, W, L)$. Triaxial
ellipsoids have $T < W < L$, oblate and prolate spheroids have $T < W = L$ and
$T = W < L$, respectively, and $L\to\infty$ for infinite elliptical cylinders.
We first fix the shape of the ellipsoid and only vary the cut's orientation
and position, and then consider randomly distributed aspect ratios $W/T$ and
$L/T$. For the sample of the random cuts thus obtained, we compute the
probability density function (PDF) of the filamentarity ${\mathcal P}(F)$.

The ellipsoids approximate the isosurfaces of a random field obtained at a
certain level, that can be varied to assess the stability of the results and
thus to isolate physically significant features of the isosurfaces. This is
done in Section~\ref{GT}. We stress, however, that there is no need to
approximate the structures with ellipsoids in applications since the
filamentarity of shapes seen in a 2D cross-section can be computed directly
using Eq.~\eqref{eq:F}.

\begin{figure}
\centering
\includegraphics[width=\columnwidth]{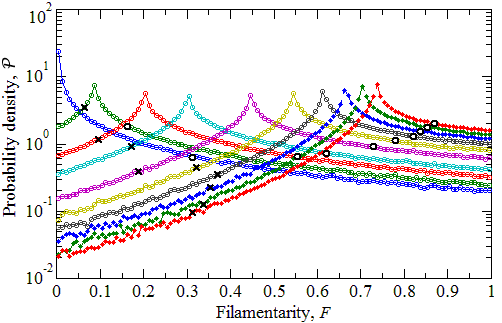}\\
\caption{\label{fig:Inf}The probability density of the filamentarity
${\mathcal P}(F)$ of random cuts of infinitely long elliptical cylinders with
the base aspect ratios $W/T = 1, 2, 3, 4, 6, 8, \dots, 16$ (different values
of $W/T$ are identified with colour as in Fig.~\ref{fig:FinCrosSecs}),
obtained from about 400,000 individual cuts binned into $120$ intervals in
$F$. The black markers on each curve indicate the ranges where the asymptotics
at $F\to0$ and $F\to1$ are accurate to within 10\%.
}
\end{figure}
\begin{figure}
\centering
\includegraphics[width=\columnwidth]{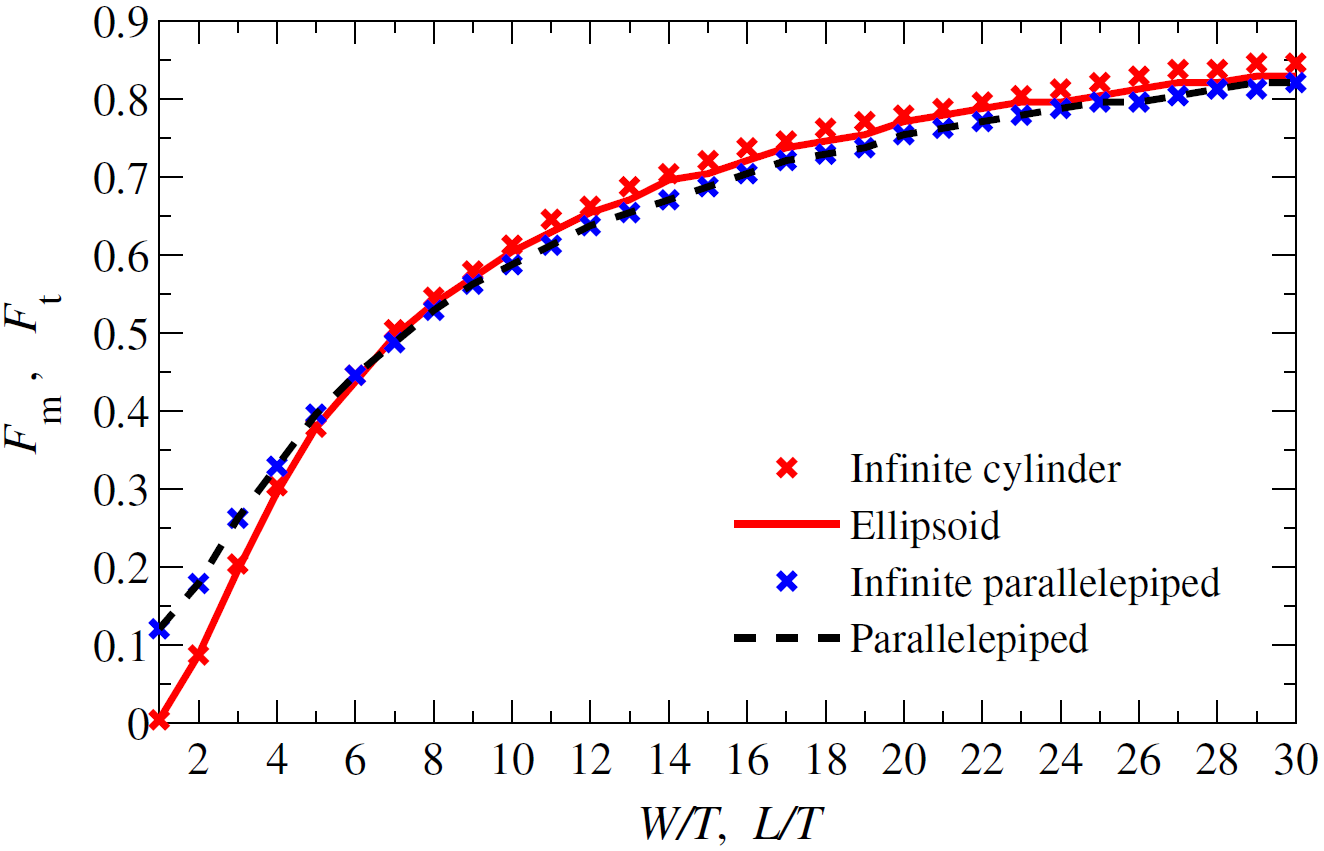}
\caption{\label{InFm}The modal filamentarity $F\mx$
versus $W/T$ and the truncation filamentarity $F\trunc$ versus $L/T$. $F\mx$
is the value of $F$ where $\mathcal{P}(F)$ is maximum, and  $F\trunc$ where
$\mathcal{P}(F)$ is truncated for finite-length structures. Results for
elliptical and rectangular-base cylinders of finite length are shown with
curves (solid/red and dashed/blue, respectively), those for shapes of infinite
length are shown with crosses. The curves representing $F\mx$ and $F\trunc$
for similar shapes overlap perfectly. The difference between the curves near
the origin is due to the difference between a circle and a square.
}
\end{figure}

\section{\label{sec:Inf}Ellipsoids, from filaments to pancakes}
We start with the simplest case of ideal filaments represented by a sample of
randomly oriented, infinitely long elliptical cylinders with the base axis
lengths $W$ and $T$. The probability density of the filamentarity in a
randomly chosen cross-section of the sample, ${\mathcal P}(F)$, is shown in
Fig.~\ref{fig:Inf} for various values of $W/T$. The PDF has a single sharp,
cuspy maximum at a certain $F=F\mx$ (called the \textit{modal\/} value of $F$)
that depends on $x=W/T$ as in Eq.~(\ref{Fx}) or Fig.~\ref{InFm} for $1\leq
x<\infty$ since $W/T\geq1$ by definition.

The filamentarity PDF has distinct asymptotic behaviours, also sensitive to
$W/T$: exponential for $F\to0$ and a power law at $F\to1$. We do not present
the corresponding fits here as they may be model dependent. We also considered
infinite and finite cylinders with a rectangular base to obtain similar
results but with $\mathcal{P}=0$ for $F<0.12$.

\begin{figure}
\centering
\includegraphics[width=0.995\columnwidth]{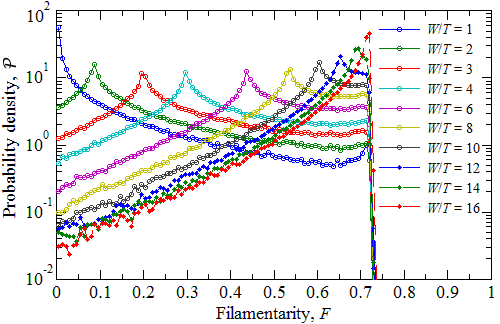}
\caption{\label{fig:FinCrosSecs}As Fig.~\ref{fig:Inf}, but for 200,000 random
cuts of triaxial ellipsoids with $L/T = 16$.
}
\end{figure}

The effect of a finite length of the structure should be the strongest when
the cross-section is nearly aligned with the structure's major axis, i.e., at
large $F$. This expectation is corroborated by Fig.~\ref{fig:FinCrosSecs}
obtained for triaxial ellipsoids with fixed $L/T=16$ and various widths,
$1\leq W/T\leq16$. Unlike the case of an infinite cylinder shown in
Fig.~\ref{fig:Inf}, the PDF of Fig.~\ref{fig:FinCrosSecs} is truncated:
${\mathcal P}=0$ at $F>F\trunc\approx0.73$ (the filamentarity of an ellipse
with the axis ratio 16). The truncation filamentarity $F\trunc$ provides a
direct measure of $L/T$.

Comparison of Figs~\ref{fig:Inf} and \ref{fig:FinCrosSecs} shows that infinite
cylinders and triaxial ellipsoids with the same $W/T$ have (nearly) the same
modal filamentarity $F\mx$, so that Eq.~(\ref{Fx}) or Fig.~\ref{InFm} can be
used to obtain $W/T$ from $F\mx$ in either case. Remarkably, the dependence of
$F\trunc$ on $L/T$ is identical to that of $F\mx$ on $W/T$, and both aspect
ratios can be recovered from Eq.~(\ref{Fx}), or a similar equation for the
ellipse, as it merely relates the relevant aspect ratio to the filamentarity.

The two most prominent aspects of the filamentarity PDF, its maximum and
truncation, obtained from a single 2D cross-section provide complete, easily
accessible information on the shape of the structures in 3D: both $W/T$ and
$L/T$ can be recovered from $F\mx$ and $F\trunc$, respectively, using
Eq.~(\ref{Fx}) or its analogue for the ellipse.

As a circular filament ($W/T=1$, $L/T=16$) gradually turns into a pancake
($W/T=L/T=16$), $F\mx$ increases. This, perhaps counterintuitive, result can
easily be understood: a large fraction of the random cuts of the filament are
nearly circular ($F\approx 0$), whereas the majority of the pancake's cuts are
elongated.

\begin{figure}
\centering
\includegraphics[width=\columnwidth]{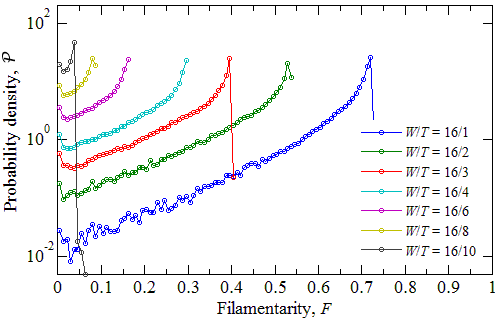}
\caption{\label{fig:Sph}The filamentarity PDF for the cuts of oblate spheroids
of $W = L = 16$ and $T = 1, 2, 3, 4, 6, 8, 10$, obtained from $\simeq72,000$
individual cuts binned into $120$ intervals in $F$. Here, $F\mx=F\trunc$.
}
\end{figure}

\section{\label{Sph}Spheroids, from pancakes to spheres}
For prolate spheroids ($W=T<L$), the PDF maximum is always at $F\mx=0$ (as one
of the cuts is circular), so the left tail of $\mathcal{P}(F)$ disappears
($W/T=1$ in Fig.~\ref{fig:FinCrosSecs}), but $F\trunc$ remains the same
function of $L/T$. For oblate spheroids ($T<W=L$), the modal and truncation
filamentarities are the same, $F\mx=F\trunc$, as shown with filled red circles
($W/T=16$) in Fig.~\ref{fig:FinCrosSecs} and, separately, in
Fig.~\ref{fig:Sph}. As an oblate spheroid (pancake) gradually turns into a
sphere, the maximum of the PDF shifts to the left, $F\mx\to0$. The value of
$F\mx$ immediately gives us the value of $W/T = L/T$ using Eq.~(\ref{Fx}).

\begin{figure}
\centering
\includegraphics[width=\columnwidth]{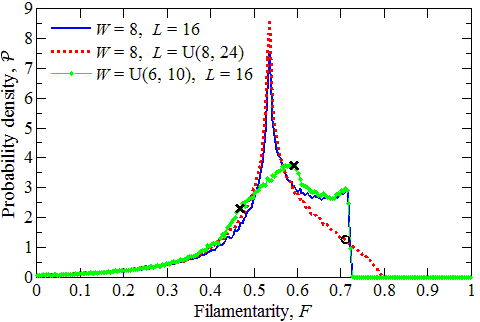}
\caption{\label{fig:EldLd}The filamentarity PDF for the cuts of triaxial
ellipsoids of fixed shape $W/T = 8$, $L/T = 16$ (blue solid line); ellipsoids
with a range of lengths, $W/T = 8$, $L/T\sim U(8,24)$ (red dots); and
ellipsoids with a range of widths, $L/T = 16$, $W/T\sim U(6,10)$ (green dots)
(288,000 realizations with $240$ bins in $F$).
}
\end{figure}

\begin{figure*}
\centering
\includegraphics[width=0.99\textwidth]{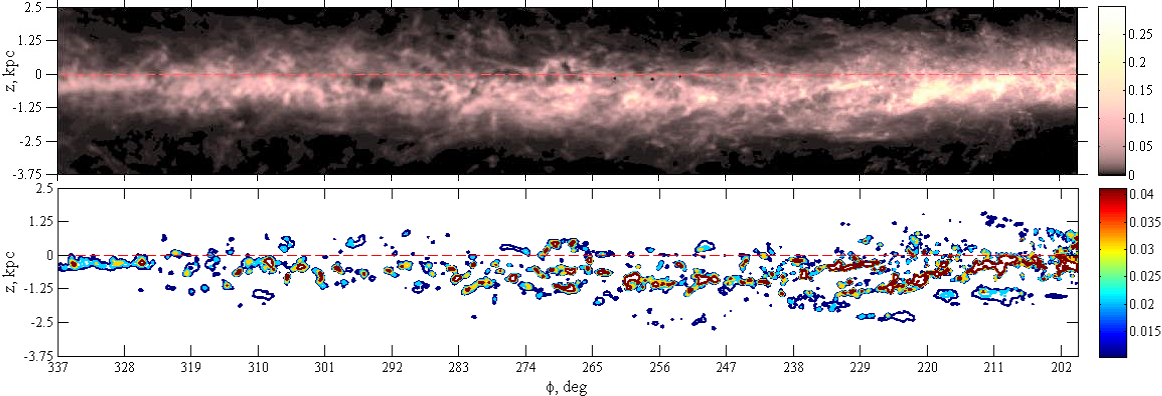}
\caption{\label{HImap}\textit{Upper panel:\/} the H\,{\sc i} number density
(in cm$^{-3}$) from the GASS survey \citep{GASS10} at $r=16\,$kpc,
$200^\circ\leq\phi\leq337^\circ$, $-3.75<z<2.5\kpc$, with $(r,\phi,z)$ the
Galactocentric cylindrical coordinates with the Sun at
$(r,\phi,z)=(8.5\,\mathrm{kpc},180^\circ,0$). \textit{Lower panel:\/} The
isocontours of the fluctuations $n$ in the gas number density obtained from
the data of the upper panel. The mean value of $n$ is zero; the contours shown
are at $n=\nu\sigma$, with $\nu= 1, 2, 3, 4, 5$ and
$\sigma=0.01\,\mathrm{cm}^{-3}$ the r.m.s.\ value of $n$ in this region.
}
\end{figure*}

\begin{figure}
\centering
\includegraphics[width=\columnwidth]{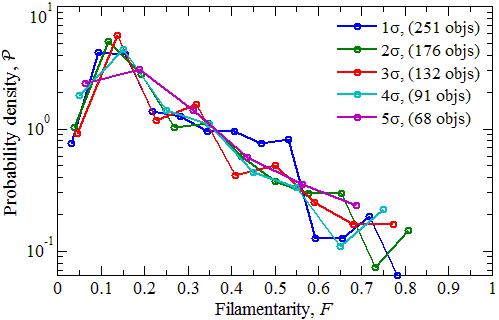}
\caption{\label{fig:Obs}$\mathcal{P}(F)$ for a 2D cross-section of the
fluctuations of neutral hydrogen density $n$ in the Milky Way at a
Galactocentric radius $R = 16 \kpc$. Different curves refer to various levels,
$n=\nu\sigma$ with $\sigma$ the r.m.s.\ value of $n$ and $\nu=1,2,\ldots5$.
}
\end{figure}

\section{\label{Spr}A range of shapes}
Now, consider how $\mathcal{P}(F)$ changes when the ellipsoids have randomly
distributed aspect ratios, conservatively chosen using uniform probability
distributions. More localized distributions would have a weaker effect.

Figure~\ref{fig:EldLd} shows, with red filled circles, ${\mathcal P}(F)$ for
ellipsoids of $W/T = 8$ and $L/T$ uniformly distributed in the interval $L_1
\leq L \leq L_2$, where $L_1/T = 8$ and $L_2/T = 24$, i.e., $L/T\sim U(8,24)$,
with the mean value $\langle L/T \rangle=16$ and the standard deviation
$\sigma_{L/T} = 4.62$. The PDF maximum is at the same value of $F\mx$ and
remains as sharp as for the fixed length $L/T=16$ (blue line in
Fig.~\ref{fig:EldLd} and yellow, in Fig.~\ref{fig:FinCrosSecs}), but the right
tail of the PDF becomes longer and has an inflection point at $F\infl=0.71$
corresponding to $\langle L/T \rangle=16.5$ (provided $W<L_1$); the inflection
point is marked with open circle. Thus, for ellipsoids with a range of
lengths, the inflection point of ${\mathcal P}(F)$ is close to the value of
$F$ where ${\mathcal P}$ is truncated for ellipsoids whose relative length is
$\langle L/T\rangle$. The position of $F\trunc$, such that ${\mathcal
P}(F\trunc)=0$, provides a direct measure of the maximum length $L_2/T$:
$F\trunc\approx0.8$ in Fig.~\ref{fig:EldLd}, corresponding to $L_2/T\approx24$
from Fig.~\ref{InFm} or Eq.~(\ref{Fx}). The extent of the right tail from the
inflection point to $F\trunc$ is proportional to $\sigma_L$:
$F\trunc-F\infl\approx 0.1\sigma_{L/T}$. Thus, Eq.~(\ref{Fx}) or a similar
equation for ellipses can be used to determine the parameters of random-shape
ellipsoids from the corresponding values of $F$.

The right tail of the PDF is slightly different when $W>L_1$: the inflection
point disappears but $\log{\mathcal P}(F)$ still has it. Prolate spheroids
with randomly distributed length ($T = W < L_1$) produce a similar picture:
the truncation point gives $L_2/T$, and the inflection point gives $\langle
L/T \rangle$.

A uniform distribution of the ellipsoid widths $W$ at fixed $T$ and $L$ has a
stronger, but easily understandable, influence on the shape of the PDF. The
peak, sensitive to $W/T$, becomes lower and broader  (the green line in
Fig.~\ref{fig:EldLd}), but the most probable value of $F$ depends on the mean
value of $W/T$ exactly as before. Its width (between the points marked with
crosses on the green curve), $0.47 \leq F\mx \leq 0.6$, reflects directly the
range $6.5 \lesssim W/T \lesssim 10.2$, using Eq.~(\ref{Fx}).

\section{Galactic turbulence}\label{GT}
The interstellar medium (ISM) of spiral galaxies is involved in transonic
turbulent motions \citep{ES04,SE04}. As a result, the observed distribution of
neutral hydrogen is dominated by filamentary structures which may be partially
due to quasi-spherical shells viewed tangentially, and it remains unclear if
the filaments are real or just an artifact of projection \citep{KH88}.
Numerical simulations of the ISM also show filamentary structures produced by
compression \citep{B-PV-SS99,KBSTN99}. However, the morphology of the gas
filaments has never been quantified in either observations or simulations.

The upper panel of Fig.~\ref{HImap} shows a cross-section of the fluctuations
in the number density of neutral hydrogen, H\,{\sc i}, in the Milky Way from
the GASS survey \citep{GASS10} at a constant Galactocentric radius of 16\,kpc.
We have subtracted large-scale trends along and across the gas layer to obtain
gas density fluctuations $n$ with vanishing mean value shown in the lower
panel (Makarenko et al., in preparation).

The lower panel of Fig.~\ref{HImap} shows the isocontours of $n$. To ensure
that contours obtained at various levels are not strongly correlated, they are
separated in $n$ by its standard deviation $\sigma$, $n=\nu\sigma$ with
integer $\nu$. The filamentarity of individual contours in Fig.~\ref{HImap}
was obtained using Eq.~\eqref{eq:F} from their enclosed areas and perimeters
measured directly in the image (essentially by counting pixels within a
contour and on its boundary), \textit{without\/} approximating them as
ellipses (or any other shapes).

The PDFs of the filamentarity of the contours of Fig.~\ref{HImap} are shown in
Fig~\ref{fig:Obs}. The PDFs at all levels are remarkably similar. This
suggests a self-similar shape of the isodensity surfaces in this range of
$\nu$. The PDFs are all similar to those of triaxial ellipsoids as they have a
well pronounced maximum (note the logarithmic scale of the vertical axis), a
power-law tail at larger $F$ and a clear truncation that occurs at the same
$F$ for all the isosurface levels; in particular, none of about 700 structures
discernible in the lower panel of Fig.~\ref{HImap} has $F$ exceeding
approximately 0.8. The form of the PDF indicates that the H\,{\sc i}
distribution is indeed filamentary in 3D (compare Figs~\ref{fig:Obs} and
\ref{fig:FinCrosSecs}). The locations of the maximum at $F\mx=0.10$--0.15 and
truncation at $F\trunc=0.7$--0.8 yield, using Eq.~(\ref{Fx}), the mean width
of the filaments as $W/T=2$--3 and their largest length as
$L_2/T\approx15$--25.

The scatter of the curves in Fig.~\ref{fig:Obs} provides a measure of the
errors involved. For example, the most probable value of $F$ varies in the
range $0.09<F\mx<0.16$ for $1\geq\nu\geq4$, leading to an uncertainty of
$\pm0.3$ in $W/T$. The value of $F\trunc$ might be affected by insufficient
statistics as it is obtained from the low-probability tail of
$\mathcal{P}(F)$. Although $\mathcal{P}(F\trunc)\simeq0.1\text{--}0.2$ is
reasonably high, further analysis is required. The accuracy of the results can
be improved by combining several cross-sections but we leave this to a
detailed analysis of the H\,{\sc i} density fluctuations which will be
published elsewhere.

\section{What can be learned from the filamentarity PDF?}
The probability density function $\mathcal P$ of the filamentarity $F$ of the
individual structures in a 2D cross-section of a 3D sample of isotropically
oriented triaxial ellipsoids is a sensitive diagnostic of the 3D morphology of
the ellipsoids.

\renewcommand{\theenumi}{\roman{enumi}}
\renewcommand{\labelenumi}{(\theenumi)}

If all the ellipsoids have the same axis ratios $W/T$ and $L/T$ (but not
necessarily the same size):
\begin{enumerate}
\item ${\mathcal P}(F)$ has a single sharp maximum at $F=F\mx$, and $F\mx$ is
related uniquely and simply to $x=W/T$ as given by Eq.~(\ref{Fx}) and shown in
Fig.~\ref{InFm}.

\item For structures of finite length, ${\mathcal P}=0$ for $F>F\trunc$, and
$F\trunc$ depends on $L/T$ exactly as $F\mx$ depends on $W/T$.

\item Non-zero probability at $F\to0$ indicates that the ellipsoids have a
circular cross-section.

\item Non-zero probability at $F\to1$ indicates that the length of the
ellipsoids is comparable to or exceeds the size of the volume sampled, so that
they are effectively infinitely long.

\item When $F\mx=F\trunc$, the structures are oblate spheroids. The closer
$F\mx$ is to zero, the closer the objects are to spheres.

\item When $F\mx = 0$, the structures are prolate spheroids.
\end{enumerate}
We have also considered the effect of randomly distributed $W/T$ and $L/T$,
i.e., random structures of variable shape. For the width and length uniformly
distributed over the ranges $W_1<W<W_2$ and $L_1<L<L_2$, we have shown that:
\begin{enumerate}
\item The maximum of ${\mathcal P}$ at $F=F\mx$ becomes
flatter and loses a cusp, but its position depends on the mean value of $W$
exactly as above. The width of the maximum is related to the range of $W$.

\item The abrupt truncation of ${\mathcal P}$ at larger $F$ is replaced by a
range of $F$ where ${\mathcal P}$ decreases smoothly before being truncated.
Now, ${\mathcal P}(F)$ acquires an inflection point at $F=F\infl$
corresponding to the mean value of $L/T$. The eventual truncation to
${\mathcal P}=0$ occurs at $F=F\trunc$ that uniquely depends on $L_2/T$, and
$F\trunc-F\infl$ is proportional to $L_2-L_1$.
\end{enumerate}

Altogether, the PDF of the filamentarity in a 2D cross-section provides
remarkably rich and easily accessible information about the morphology of the
3D random structures which can be recovered from a single, simple dependence
of Eq.~(\ref{Fx}) or its analogue for ellipses. Anisotropic structure
distributions produce a secondary maximum in ${\mathcal P}(F)$, which can be
used to characterize the anisotropy as will be discussed elsewhere.

We have demonstrated the efficiency of this approach to morphological analysis
using isosurfaces of the density fluctuations in the compressible, turbulent
interstellar gas in the Milky Way. We have confirmed that the structures
visible in a 2D cross-section of the isosurfaces are consistent with
isotropically distributed filaments (rather than shells or other flattened
objects) and estimated, for the first time, the relative width and length of
the filamentary structures.

\mbox{}\vspace*{-27pt}

\section*{Acknowledgments}
We are grateful to Rodion Stepanov and Thomas Buchert for helpful comments.
Thoughtful suggestions of the referee, Jeffrey Scargle, have helped to improve
the text. Useful discussions with Iioannis Ivrissimtzis,  Shiping Liu, Norbert
Peyerimhoff and Alina Vdovina are gratefully acknowledged. This work was
supported by the Leverhulme Trust (grant RPG-097) and STFC (grant
ST/L005549/1).

\bibliographystyle{mn2e}
\bibliography{mfs}

\begin{thebibliography}{}

\bibitem[\protect\citeauthoryear{{Adler} \& {Taylor}}{{Adler} \&
  {Taylor}}{2007}]{AT07}
{Adler} R.~J.,  {Taylor} J.~E.,  2007, Random Fields and Geometry.
Springer, New York

\bibitem[\protect\citeauthoryear{{Ballesteros-Paredes}, {V{\'a}zquez-Semadeni}
  \& {Scalo}}{{Ballesteros-Paredes} et~al.}{1999}]{B-PV-SS99}
{Ballesteros-Paredes} J.,  {V{\'a}zquez-Semadeni} E.,    {Scalo} J.,  1999,
  \apj, 515, 286

\bibitem[\protect\citeauthoryear{{Elmegreen} \& {Scalo}}{{Elmegreen} \&
  {Scalo}}{2004}]{ES04}
{Elmegreen} B.~G.,  {Scalo} J.,  2004, \araa, 42, 211

\bibitem[\protect\citeauthoryear{{Kalberla}, {McClure-Griffiths}, {Pisano},
  {Calabretta}, {Ford}, {Lockman}, {Staveley-Smith}, {Kerp}, {Winkel}, {Murphy}
  \& {Newton-McGee}}{{Kalberla} et~al.}{2010}]{GASS10}
{Kalberla} P.~M.~W.,  {McClure-Griffiths} N.~M.,  {Pisano} D.~J.,  {Calabretta}
  M.~R.,  {Ford} H.~A.,  {Lockman} F.~J.,  {Staveley-Smith} L.,  {Kerp} J.,
  {Winkel} B.,  {Murphy} T.,    {Newton-McGee} K.,  2010, \aap, 521, A17

\bibitem[\protect\citeauthoryear{{Korpi}, {Brandenburg}, {Shukurov}, {Tuominen}
  \& {Nordlund}}{{Korpi} et~al.}{1999}]{KBSTN99}
{Korpi} M.~J.,  {Brandenburg} A.,  {Shukurov} A.,  {Tuominen} I.,    {Nordlund}
  {\AA}.,  1999, \apjl, 514, L99

\bibitem[\protect\citeauthoryear{{Kulkarni} \& {Heiles}}{{Kulkarni} \&
  {Heiles}}{1988}]{KH88}
{Kulkarni} S.~R.,  {Heiles} C.,  1988, in {Kellermann} K.~I.,  {Verschuur}
  G.~L.,  eds, Galactic and Extragalactic Radio Astronomy. Springer, New York,
  p.~95

\bibitem[\protect\citeauthoryear{{Mecke}, {Buchert} \& {Wagner}}{{Mecke}
  et~al.}{1994}]{Mecke94}
{Mecke} K.~R.,  {Buchert} T.,    {Wagner} H.,  1994, \aap, 288, 697

\bibitem[\protect\citeauthoryear{{Platz{\"o}der} \& {Buchert}}{{Platz{\"o}der}
  \& {Buchert}}{1996}]{PB96}
{Platz{\"o}der} M.,  {Buchert} T.,  1996, in {Weiss} A.,  {Raffelt} G.,
  {Hillebrandt} W.,  {von Feilitzsch} F.,   {Buchert} T.,  eds, Proc. 1st SFB
  Workshop, Astro-Particle Physics. Technische Universit\"at, M\"unich, p.~251

\bibitem[\protect\citeauthoryear{Rivollier, Debayle \& Pinoli}{Rivollier
  et~al.}{2010}]{RDP10a}
Rivollier S.,  Debayle J.,    Pinoli J.~C.,  2010, Aust.\ J.\ Math.\ Anal.\
  Appl., 7, 1

\bibitem[\protect\citeauthoryear{{Rybicki}}{{Rybicki}}{1987}]{Ry87}
{Rybicki} G.~B.,  1987, in {de Zeeuw} P.~T.,  ed., Proc.~IAU Symp.~127,
  Structure and Dynamics of Elliptical Galaxies. D.~Reidel, Dordrecht, p.~397

\bibitem[\protect\citeauthoryear{{Sahni}, {Sathyaprakash} \&
  {Shandarin}}{{Sahni} et~al.}{1998}]{Sahni98}
{Sahni} V.,  {Sathyaprakash} B.~S.,    {Shandarin} S.~F.,  1998, \apjl, 495, L5

\bibitem[\protect\citeauthoryear{{Scalo} \& {Elmegreen}}{{Scalo} \&
  {Elmegreen}}{2004}]{SE04}
{Scalo} J.,  {Elmegreen} B.~G.,  2004, \araa, 42, 275

\bibitem[\protect\citeauthoryear{{Schmalzing}, {Buchert}, {Melott}, {Sahni},
  {Sathyaprakash} \& {Shandarin}}{{Schmalzing} et~al.}{1999}]{SBMSSS99}
{Schmalzing} J.,  {Buchert} T.,  {Melott} A.~L.,  {Sahni} V.,  {Sathyaprakash}
  B.~S.,    {Shandarin} S.~F.,  1999, \apj, 526, 568

\bibitem[\protect\citeauthoryear{Schneider \& Weil}{Schneider \&
  Weil}{2008}]{SW08}
Schneider R.,  Weil W.,  2008, Stochastic and Integral Geometry.
Springer, Berlin

\bibitem[\protect\citeauthoryear{{Vinci}, {Freeman}, {Newman}, {Wasserman} \&
  {Genovese}}{{Vinci} et~al.}{2014}]{VFNWG14}
{Vinci} G.,  {Freeman} P.,  {Newman} J.,  {Wasserman} L.,    {Genovese} C.,
  2014, in Statistical Challenges in the 21st Century Cosmology, IAU Symp.~306.
Cambridge Univ. Press, Cambridge (in press; arXiv: 1406.7536)

\bibitem[\protect\citeauthoryear{{Wiegand}, {Buchert} \& {Ostermann}}{{Wiegand}
  et~al.}{2014}]{WBO14}
{Wiegand} A.,  {Buchert} T.,    {Ostermann} M.,  2014, \mnras, 443, 241

\end{thebibliography}
\label{lastpage}
\end{document}